\documentclass[letter,tradiabstract]{aa}

\usepackage{natbib}
\usepackage[varg]{txfonts}
\usepackage{graphicx}
\usepackage{url}

\bibpunct{(}{)}{;}{a}{}{,}

\usepackage{color,soul}

\usepackage[utf8]{inputenc}

\begin{document}

\title{First results from the CALYPSO IRAM-PdBI survey\thanks{Based on
    observations carried out with the IRAM Plateau de Bure
    interferometer. IRAM is supported by INSU/CNRS (France), MPG
    (Germany), and IGN (Spain).}\,\thanks{Table \ref{tab:results} is
    only available in electronic form via http://www.edpsciences.org}}

\subtitle{I. Kinematics of the inner envelope of \object{NGC1333-IRAS2A}}

\titlerunning{Kinematics of the inner envelope of NGC1333-IRAS2A}

\author{S.~Maret \inst{1} \and A.~Belloche \inst{2} \and A.~J. Maury
  \inst{3} \and F.~Gueth \inst{4} \and Ph.~André \inst{5} \and
  S.~Cabrit \inst{6,1} \and C.~Codella \inst{7} \and S.~Bontemps \inst{8,9}}

\institute{UJF-Grenoble 1 / CNRS-INSU, Institut de Planétologie et
  d'Astrophysique de Grenoble, UMR 5274, Grenoble, F-38041, France
  \and Max-Planck-Institut für Radioastronomie, Auf dem Hügel 69,
  53121 Bonn, Germany \and Harvard-Smithsonian Center for
  Astrophysics, 60 Garden street, Cambridge, MA 02138, USA \and IRAM,
  300 rue de la piscine, 38406 Saint Martin d'Hères, France \and
  Laboratoire AIM-Paris-Saclay, CEA/DSM/Irfu - CNRS - Université Paris
  Diderot, CE-Saclay, F-91191 Gif-sur-Yvette, France \and LERMA,
  Observatoire de Paris, CNRS, ENS, UPMC, UCP, 61 Av de
  l'Observatoire, F-75014 Paris, France \and INAF-Osservatorio
  Astrofisico di Arcetri, Largo E. Fermi 5, I-50125 Firenze, Italy
  \and Université de Bordeaux, LAB, UMR 5804, F-33270 Floirac, France
  \and CNRS, LAB, UMR 5804, F-33270 Floirac, France}

\date{Received ...; accepted ...}
 
\abstract{The structure and kinematics of Class 0 protostars on scales
  of a few hundred AU is poorly known. Recent observations have
  revealed the presence of Keplerian disks with a diameter of
  150-180~AU in \object{L1527-IRS} and \object{VLA1623A}, but it is
  not clear if such disks are common in Class 0 protostars. Here we
  present high-angular-resolution observations of two methanol lines
  in NGC1333-IRAS2A. We argue that these lines probe the inner
  envelope, and we use them to study the kinematics of this
  region. Our observations suggest the presence of a marginal velocity
  gradient normal to the direction of the outflow. However, the
  position velocity diagrams along the gradient direction appear
  inconsistent with a Keplerian disk. Instead, we suggest that the
  emission originates from the infalling and perhaps slowly rotating
  envelope, around a central protostar of
  $0.1-0.2~\mathrm{M_\sun}$. If a disk is present, it is smaller than
  the disk of L1527-IRS, perhaps suggesting that NGC1333-IRAS2A is
  younger.}

\keywords{ISM individual objects: NGC~1333-IRAS~2A -- ISM: kinematics
  and dynamics -- ISM: molecules -- Stars: formation}

\maketitle

\section{Introduction}

Understanding the first steps of the formation of protostars and
protoplanetary disks is a major unsolved problem of modern
astrophysics. Observationally, the key to constraining protostar
formation models lies in high-resolution studies of the youngest
protostars. Because their lifetime is only $t \sim 10^{5}$~yr and most
of their mass is still in the form of a dense envelope ($M_\mathrm{*}
<< M_\mathrm{env}$ ), Class 0 protostars are likely to retain detailed
information on the initial conditions and detailed physics of the
collapse phase \citep{Andre00}. However, the physics of Class 0
protostars is still surprisingly poorly understood, due to the
paucity of the sub-arcsecond (sub)mm observations required to probe their
innermost (100 AU) regions. Several basic questions thus remain
open, such as the mere existence of accretion disks during the
Class 0 phase, or the structure of the velocity field in Class 0
envelopes (e.g., relative importance of inflow, rotation and outflow). On
the large scales, some of the Class 0 sources show evidence of
flattened structures, extending roughly perpendicular to the outflow
over $\ge 10,000$~AU \citep{Belloche02,Looney07,Tobin10}. However,
line observations indicate that these structures are slowly
rotating infalling envelopes rather than centrifugally
supported disks \citep{Belloche02,Chiang10}. On smaller scales
(several hundred AU), little is known on the envelope/disk structure
and kinematics. Only two Class 0 protostars, L1527-IRS
and VLA1623A, show an evidence of a compact (150 AU to 180 AU in
diameter) Keplerian disk \citep{Tobin12b,Murillo13}.  Studying these
regions requires observations of specific (optically thin) tracers of
the inner region to avoid contamination by the outflow and the
extended envelope. Methanol lines are good specific probes of these
regions, as we discuss below.
  
In this letter, we present observations of two methanol lines in the
NGC1333-IRAS2A Class 0 protostar (hereafter IRAS2A) located in the
Perseus molecular complex at 235~pc \citep{Hirota08}.  These
observations are part of the CALYPSO (Continuum And Line in Young
Proto-Stellar
Objects) survey\footnote{\url{http://irfu.cea.fr/Projets/Calypso/}},
an IRAM Plateau de Bure Interferometer (hereafter PdBI) large program
that aims at studying a large sample of Class 0 protostars at
sub-arcsecond resolution. We use these lines to probe the kinematics
of the protostellar envelope on $\sim$200~AU scales, and to determine
if a disk is present. Two companion letters present the complex
organic molecule (COM) emission \citep{Maury13} and the jet properties
\citep{Codella13}.

\section{Observations}
\label{sec:observations}

\begin{figure*}
  \centering
  \includegraphics[width=17cm]{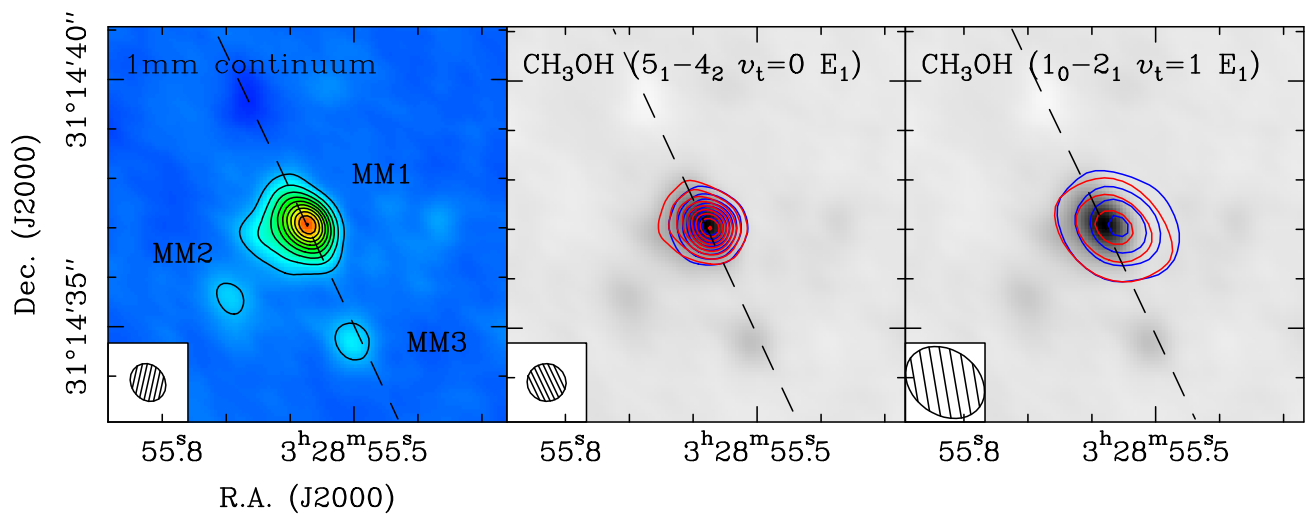}
  \caption{Continuum and line emissions observed towards IRAS2A with
    the PdBI. The left panel shows the 1 mm continuum
    emission. Contour spacing is 10~mJy~beam$^{-1}$ (6.6$\sigma$).
    The other panels show the velocity integrated intensities of the
    1~mm and 3~mm methanol lines (blue and red contours) together with
    the continuum (grayscale image). The blue and red contours
    correspond to velocities $< 6.5$~km~s$^{-1}$ and $>
    6.5$~km~s$^{-1}$, respectively. In the center panel, the blue and
    red contours are almost coincident.  The contour spacings are
    100~mJy~beam$^{-1}$~km~s$^{-1}$ (3.2$\sigma$) and
    30~mJy~beam$^{-1}$~km~s$^{-1}$ (2.1$\sigma$) for the 1~mm and 3~mm
    lines, respectively. In each panel the dashed ellipse indicates
    the size and orientation of the synthesized beam. The dashed line
    shows the orientation of the outflow \citep[which has a P.A. of
    25\degr; ][]{Codella13}.}
  \label{fig:maps}
\end{figure*}

\begin{figure*}
  \sidecaption
  \includegraphics[width=12cm]{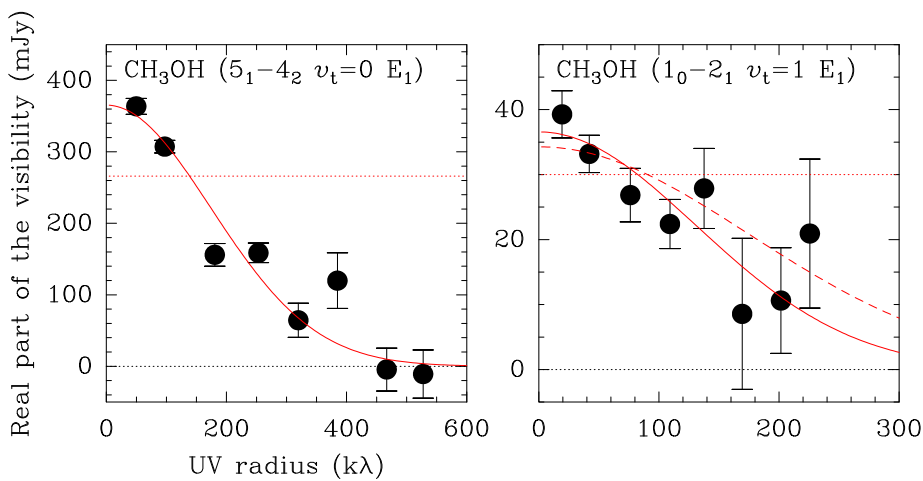}
  \caption{Real part of the visibility of the 1~mm (left panel) and
    3~mm (right panel) methanol lines as a function of the UV radius
    (black points with error bars). The visibilities have been
    averaged over the line profile, and then averaged circularly. The
    red solid line in each panel shows the result of a Gaussian source
    fit, while the red dotted line shows the result of a point source
    fit. The dashed line in the right panel shows the result of a
    Gaussian source fit with a FWHM size fixed to that of the 1~mm
    line (0.44$\arcsec$). The black dotted line in each panel
    indicates the zero level.}
  \label{fig:uv-fit}
\end{figure*}

Observations of IRAS2A were carried out with the PdBI between November
2010 and February 2011 using the A and C configurations of the
array. The
$\mathrm{CH_{3}OH}~(5_{1}-4_{2}~\varv_{\mathrm{t}}=0~\mathrm{E}_{1})$
line at 216.945\,600~GHz (1~mm) was observed using the narrow-band
backend, providing a bandwidth of 512 channels of 39~kHz
(0.05~km~s$^{-1}$) each. The
$\mathrm{CH_{3}OH}~(1_{0}-2_{1}~\varv_{\mathrm{t}}=1~\mathrm{E}_{1})$
line at 93.196\,670~GHz (3~mm) was also observed using the narrow-band
backend, but with 256 channels of 312~kHz (1.0~km~s$^{-1}$)
each. Calibration was done using {\tt CLIC}, which is part of the {\tt
  GILDAS} software
suite\footnote{\url{http://www.iram.fr/IRAMFR/GILDAS/}}. For the 1~mm
observations, the phase RMS was $< 80\degr$, with precipitable water
vapor (PWV) between 0.5~mm and 2~mm, and system temperatures
($T_\mathrm{sys}$) between 100~K and 250~K. For the 3~mm observations,
the phase RMS was $< 40\degr$, with PWV of 2-3~mm and $T_\mathrm{sys}$
of 70-80~K. The continuum emission was removed from the visibility
tables to produce continuum-free line tables. The visibilities of the
1~mm methanol line were resampled at a spectral resolution of
$0.5$~km~s$^{-1}$ to improve the signal-to-noise ratio.  Spectral
datacubes were produced from the visibility tables using a natural
weighting, and deconvolved using the standard CLEAN algorithm in the
{\tt MAPPING} program. The synthesized beam is $0.79\arcsec \times
0.76\arcsec$ (P.A. $70\degr$) and $1.71\arcsec \times 1.31\arcsec$
(P.A. $54\degr$) for the 1~mm and 3~mm lines, respectively. The RMS
noise per channel in the final datacubes is 21~mJy/beam and 5~mJy/beam
for the 1~mm and 3~mm lines, respectively.

\section{Results and analysis}
\label{sec:results-analysis}

\begin{figure*}
  \sidecaption
  \includegraphics[width=12cm]{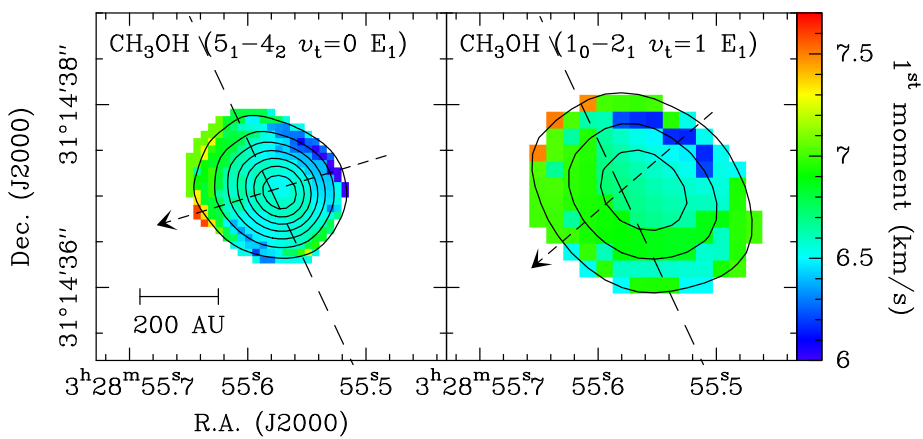}
  \caption{First-order moment maps of the 1~mm (left panel) and 3~mm
    (right panel) methanol lines. In each panel the solid contours
    show the velocity integrated line intensity. The contour spacings
    are 200~mJy~beam$^{-1}$~km~s$^{-1}$ and
    20~mJy~beam$^{-1}$~km~s$^{-1}$ for the 1~mm and 3~mm methanol
    lines, respectively.  The dashed line shows the orientation of the
    outflow, while the dashed arrow indicates the orientation of the
    velocity gradient as determined from the first-order moment fit
    (see Sect.~\ref{sec:results-analysis}). The line and arrow
    intersect at the position of MM1.}
  \label{fig:moment1}
\end{figure*}

Figure~\ref{fig:maps} shows the 1~mm continuum emission together with
velocity integrated intensity maps of the 1~mm and 3~mm methanol
lines. Continuum emission is detected in three sources, which are
labeled MM1, MM2, and MM3 in \citet{Codella13}. We detect 1~mm and 3~mm
methanol line emission towards the brightest continuum peak (MM1) with
a signal-to-noise ratio at the line peaks of 21 and 8 for the 1~mm and
3~mm methanol lines, respectively. Emission from these lines is
compact. In addition, no clear difference between the blue-shifted and
red-shifted emission (at velocities $< 6.5$~km~s$^{-1}$ and $>
6.5$~km~s$^{-1}$, respectively\footnote{In the present study, we adopt
  a $v_\mathrm{LSR}$ of 6.5~km~s$^{-1}$, as determined from a fit of
  the first-order moment map of the 1~mm methanol line (see below).})
is seen.

\onltab{
  \begin{table}
    \caption{Line spectroscopic parameters, results of the visibility
      fits, and results of the first-order moment fits.}
    \label{tab:results}    
    \centering
    \begin{tabular}{l c c}
      \hline
      \hline
      Line & $5_{1}-4_{2}~\varv_{\mathrm{t}}=0~\mathrm{E}_{1}$ &
      $1_{0}-2_{1}~\varv_{\mathrm{t}}=1~\mathrm{E}_{1}$\\
      \hline
      $\nu$\tablefootmark{a} (GHz) & 216.945\,600 & 93.196\,670 \\
      $E_\mathrm{up}$ (K) & 55.9 & 309.9 \\
      \hline
      R.A. offset\tablefootmark{b} (\arcsec) & $-0.06 \pm 0.02$ & $-0.27
      \pm 0.05$ \\
      Dec. offset (\arcsec) & $0.00 \pm  0.02$ & $0.06 \pm  0.05$\\
      Size\tablefootmark{c} (\arcsec) & $0.44 \pm 0.03$ & $0.59 \pm 0.11$\\
      \hline
      $I_\mathrm{peak}$\tablefootmark{d} (mJy~beam$^{-1}$) & $457 \pm
      21$ & $24 \pm 5$ \\
      Linewidth\tablefootmark{e} (km~s$^{-1}$) & $3.30 \pm 0.04$ &
      $4.10 \pm 0.07$\\
      $\int I~\mathrm{d}v$~\tablefootmark{f}
      (Jy~beam$^{-1}$~km~s$^{-1}$) & $1.61 \pm 0.02$ & $0.11 \pm
      0.01$\\
      \hline
      $v_\mathrm{0}$\tablefootmark{g} (km~s$^{-1}$) & $6.54 \pm 0.03$ &
      $6.76 \pm 0.15$\\
      $\alpha$\tablefootmark{h} ($\degr$) & $107 \pm
      27$ & $131 \pm 36$ \\
      $G$\tablefootmark{i} (km~s$^{-1}/\arcsec$) & $0.23 \pm 0.11$ &
      $0.47 \pm 0.29$\\
      \hline
    \end{tabular}
    \tablefoot{
      \tablefoottext{a}{Frequencies are from the CDMS catalog \citep{Muller01}}
      \tablefoottext{b}{Position of the emission peak, as determined
        from the visibility fit. Offsets are relative to the
        continuum position of MM1 \citep[$03\mathrm{h} 28^\mathrm{m}
        55 \fs 58; 31\degr 14\arcmin 37\farcs057;
        \mathrm{J2000}$; ][]{Codella13}}
      \tablefoottext{c}{Deconvolved source FWHM size, assuming a circular Gaussian
        brightness.}
      \tablefoottext{d}{Line peak intensity at the position of 1~mm
        line emission peak.}
      \tablefoottext{e}{Line FWHM.}
      \tablefoottext{f}{Line integrated intensity.}
      \tablefoottext{g}{Velocity at the position of MM1, obtained
        from a fit of the first-order moment map.}
      \tablefoottext{h}{Position angle of the velocity gradient, from
        N to E.}
      \tablefoottext{i}{Velocity gradient amplitude.}
    }
  \end{table}
}
      
To determine the size of the line emission, we fitted the visibilities
assuming that the emission follows a Gaussian distribution. We
averaged the visibilities over the entire velocity range of the line,
and we performed a fit of the velocity averaged visibilities in the UV
plane. Figure~\ref{fig:uv-fit} shows the result of the fit, together
with the real part of the visibilities (averaged as a function of the
UV radius for clarity). The fit parameters are given in
Table~\ref{tab:results}. As seen in Fig.~\ref{fig:uv-fit}, the 1~mm
methanol line emission is spatially resolved, and it is reasonably
well fitted by a Gaussian source with a FWHM size of $0.44 \pm 0.03
\arcsec$. On the other hand, the 3~mm methanol line is only marginally
resolved; the emission is best fitted by a Gaussian of $0.59 \pm 0.11
\arcsec$ FWHM, but its size is also consistent (at a 2$\sigma$ level)
with that of the 1~mm line. This is also apparent in the right panel
of Fig.~\ref{fig:uv-fit}, which shows the expected real part of the
visibility for the 3~mm line, assuming a Gaussian source of
$0.44\arcsec$. The position of the 1~mm methanol line peak is, within
the uncertainties, consistent with the position of the MM1 continuum
source.  Although the position of the 3~mm methanol line appears
slightly offset, it is also consistent (at a 5$\sigma$ level) with the
position of MM1. Table~\ref{tab:results} also gives the linewidths at
the position of the 1~mm line emission peak. Both the 1~mm and 3~mm
lines have Gaussian shapes, with a FWHM of 3.3 and 4.1~km~s$^{-1}$,
respectively.

\begin{figure*}
  \sidecaption
  \includegraphics[width=12cm]{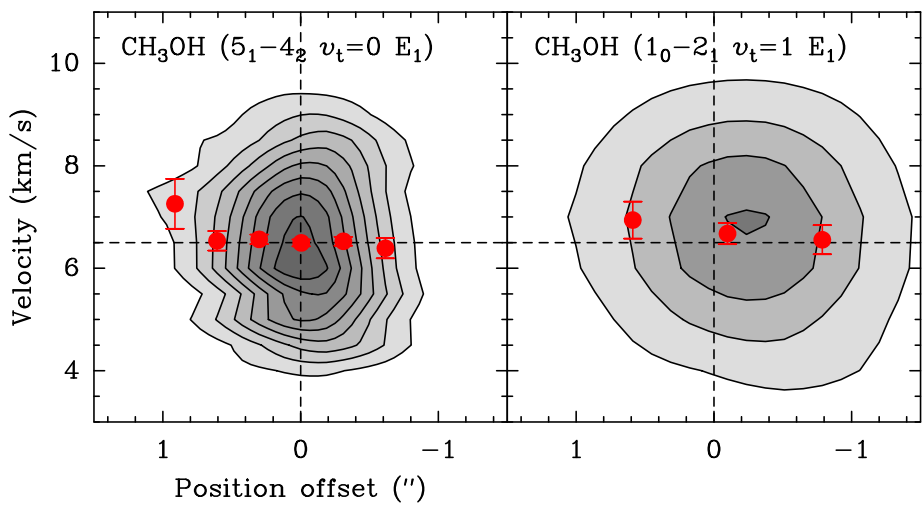}
  \caption{Position-velocity diagrams for the 1~mm (left panel) and
    3~mm (right panel) methanol lines. The contour spacing is
    50~mJy~beam$^{-1}$ (2.4$\sigma$) and 10~mJy~beam$^{-1}$
    (2$\sigma$) for the 1~mm and 3~mm methanol lines,
    respectively. The red points with error bars show the first order
    moment measured at each position offset. The horizontal dashed
    line shows the source $v_\mathrm{LSR}$ (6.5~km~s$^{-1}$). The
    vertical dashed lines indicate the center of the cut.}
  \label{fig:pv}
\end{figure*}

Figure~\ref{fig:moment1} shows the first-order moment (mean velocity)
maps for both lines. From visual inspection of the moment maps, no
clear velocity gradient is apparent. We have therefore attempted to
fit the first order moment maps with a linear function of the R.A. and
Dec. offsets. This linear gradient is expected if the emitting region
is rotating as a solid body \citep{Goodman93}.  Although solid body
rotation is not realistic, it provides a first indication of whether
or not rotation is present. Because the maps shown in
Fig.~\ref{fig:moment1} are oversampled, we have fitted the first-order
moment measured every half synthesized beam (Nyquist sampling). These
pixels were weighted by $1/\sigma^2$, where $\sigma$ is the $1
\sigma$~noise on the first-order moment, computed from the noise per
channel in the data cubes \cite[see ][Eq.~(2.3)]{Belloche13}. Results
of the fit are given in Table~\ref{tab:results}. For the 1~mm line, we
find a gradient with a P.A.  $\alpha = 107 \pm 27 \degr$ and an
amplitude $G = 0.23 \pm 0.11$~km~s$^{-1}/\arcsec$ ($202 \pm
97$~km~s$^{-1}/$pc at 235~pc). For the 3~mm, we find $\alpha = 131 \pm
36\degr$ and $G = 0.47 \pm 0.29$~km~s$^{-1}/\arcsec$ ($413 \pm
255$~km~s$^{-1}/$pc). Although the gradient detection is only marginal
(2.1$\sigma$ and 1.6$\sigma$ for the values of $G$ at 1~mm and 3~mm,
respectively), the amplitude and orientation obtained for each line
are consistent. The gradient orientations are indicated by arrows in
Fig.~\ref{fig:moment1}. Interestingly, they are roughly perpendicular
to the outflow orientation, as one would expect for rotation of the
protostellar envelope or disk.

Figure~\ref{fig:pv} shows the position-velocity (hereafter P.V.)
diagrams for both lines. The cuts were done along an axis with P.A. of
107$\degr$, and centered on the peak of the 1~mm
line. Figure~\ref{fig:pv} also shows the first-order moment measured
along the cut. No clear signature of rotation is apparent in the
P.V. diagram. Both diagrams are almost symmetric with respect to the cut
center position. However, the first-order moment measured along the
cut increases from negative to positive offsets. This is
consistent with the presence of a gradient along this axis. This
increase in the first-order moment along the cut is also marginal (see
the error bars in Fig.~\ref{fig:pv}), but the same trend is seen
for both lines.

\section{Discussion and conclusions}
\label{sec:disc-concl}

Our observations reveal compact methanol emission (0.4$\arcsec$ FWHM,
i.e., $\sim$90~AU in diameter) centered on the position of the MM1
continuum source. The observed emission has several possible
origins. First, methanol emission can arise from shocks, as a result
of ice sputtering; indeed, methanol emission is observed towards the
end points of the large scale east-west outflow of IRAS2A
\citep{Bachiller98b}. It is therefore possible that the emission
observed here is caused by shocks on smaller scales. However, the line
Gaussian shapes and small linewidths, together with the striking
contrast between the methanol and the SO and SiO emission (which are
associated with the shocked gas; see \citealt{Codella13}),
do not favor this hypothesis.

Second, the emission may arise from the inner region of the envelope
where the temperature is greater than $\sim$100~K (the so-called
hot corino). \citet{Maret05} did a single-dish survey of the methanol
emission from a sample of six Class 0 protostars, and found abundance
jumps (by about 2 orders of magnitude) in the inner envelopes of four
of them (including IRAS2A). They interpreted these abundance jumps as
due to the thermal evaporation of grain mantles in the hot
corino. Several COMs, such as acetonitrile, methyl formate, or dimethyl
ether, have also been detected in IRAS2A with single-dish
\citep{Bottinelli07} and interferometric \citep{Maury13} observations.
The similar COM \citep{Maury13} and methanol emission sizes (this
study) strongly suggest a common origin. Compact water (H$_{2}^{18}$O)
emission has also been detected in IRAS2A with the PdBI
\citep{Persson12}. The size of the H$_{2}^{18}$O emission (0.8\arcsec)
is twice as large as the methanol emission. However, the H$_{2}^{18}$O
maps show some emission associated with the outflow. This
contamination by the outflow may explain the larger extension of
the H$_{2}^{18}$O emission with respect to the methanol and the COM
emission. Dust continuum observations and modeling indicate that the
radius at which the dust temperature exceeds 100~K in IRAS2A is about
50~AU \citep{Jorgensen02}\footnote{A more recent analysis
    suggests a larger value \citep[90~AU;][]{Kristensen12}}. This
radius is in good agreement with the size of the methanol emission,
and therefore favors the hot corino scenario.

A third scenario is that the emission originates from the surface of a
disk. From SMA observations of the dust continuum, \cite{Jorgensen05b}
suggested that IRAS2A harbors a circumstellar disk of 200-300~AU in
diameter, with a mass between 0.01~M$_\sun$ and 0.1~M$_\sun$. They
suggested that the disk could be the major reservoir of the COMs and
other molecules observed in that source. \cite{Brinch09} modeled the
HCN and H$^{13}$CN (4-3) emission observed with the SMA towards IRAS2A
at 1$\arcsec$ resolution, and found that the velocity field was
dominated by infall, with little rotation. A flattened disk-like
structure dominated by inward motions is also favored by
\cite{Persson12} to explain the H$_{2}^{18}$O emission observed in
IRAS2A. Our methanol observations suggest the presence of a marginal
velocity gradient oriented roughly in a direction perpendicular to the
north-south outflow. The orientation of the gradient is similar to
that of a second east-west outflow detected in this source. However,
this second outflow is probably driven by the MM2 source and its axis
does not pass through the position of MM1 \citep{Codella13}, while the
methanol emission is clearly associated with MM1. Therefore, the
gradient is most likely caused by rotation of the inner envelope or a
disk about the outflow axis.

Regardless of the origin of the emission (i.e., hot corino or disk),
if the material traced by the methanol lines is bound to the
protostellar system (as opposed to unbound outflow material for
example), the line width can be used to estimate the dynamical mass of
the system, $M_\mathrm{dyn} \sim \left(1-2 \right) \, r \, v^{2} / G$,
where $v$ is the velocity observed at a radii $r$, and $G$ is the
gravitational constant. This expression holds if the line broadening
is due to infall motions, or Keplerian rotation about an axis close to
the plane-of-the-sky \citep[see, e.g.,][]{Terebey92}. If we assume
that the velocity at $r = 45$~AU is $1.65$~km/s (i.e., the 1~mm line
HWHM), we obtain $M_\mathrm{dyn} \simeq 0.1-0.2~\mathrm{M_\sun}$
\citep[to be compared to a total envelope mass $M_\mathrm{env} =
1.7~\mathrm{M_\sun}$;][]{Jorgensen02}. Strictly speaking,
$M_\mathrm{dyn}$ is the mass contained within a radius of 45~AU, so
that $M_{*} \lesssim 0.1-0.2~\mathrm{M_\sun}$. This estimate is in
reasonable agreement with the value of $0.25~\mathrm{M_\sun}$ obtained
by \cite{Brinch09}.

The 1~mm line P.V. diagram can be used to test the disk scenario. We
have computed synthetic P.V. diagrams for a Keplerian disk with
various values of $M_{*}$ (see Appendix~\ref{sec:synth-posit-veloc}
for details). We find that the observed P.V. diagram is inconsistent
with a Keplerian disk, regardless of the mass of the central
object. In particular, a Keplerian disk with $M_{*} =
0.1~\mathrm{M_\sun}$ would produce two peaks in the P.V. diagram,
which are not observed. On the other hand, a Keplerian disk $M_{*} =
0.01~\mathrm{M_\sun}$ would have a single peak P.V. diagram, but the
predicted linewidth would be 2-3 times smaller than observed, even for
a very turbulent disk.

To summarize, we favor a scenario in which the methanol emission
observed in IRAS2A originates in the inner envelope of the protostar,
which is infalling and perhaps slowly rotating. If a disk is present
and is emitting in methanol, it is spatially unresolved so its radius
is smaller than 45~AU (0.4\arcsec), i.e. smaller than the disk of
L1527-IRS. Since the mass of the central object in
L1527-IRS \citep[$0.2~\mathrm{M_\sun}$;][]{Tobin12b} is comparable to
that of IRAS2A, this may suggest that IRAS2A is younger.  Higher
angular resolution observations are needed to confirm the presence of
rotation in the inner parts of IRAS2A and to measure the rotation
profile precisely. In addition, observations of other Class 0
protostars are needed to obtain more statistics on the presence of
rotationaly supported disks in this phase, and to establish when disks
appear during the evolution of protostars.

\begin{acknowledgements}
  The research leading to these results has received funding from the
  European Community’s Seventh Framework Programme (/FP7/2007-2013/)
  under grant agreements No 229517 (ESO COFUND) and No 291294
  (ORISTARS), and from the French Agence Nationale de la Recherche
  (ANR), under reference ANR-12-JS05-0005.
\end{acknowledgements}

%\vspace*{-.47cm}

\bibliographystyle{aa}
\bibliography{bibliography}

\Online

\begin{appendix}

\section{Synthetic position velocity diagrams for a Keplerian
  disk}
\label{sec:synth-posit-veloc}

\begin{figure*}
  \centering
  \includegraphics[width=17cm]{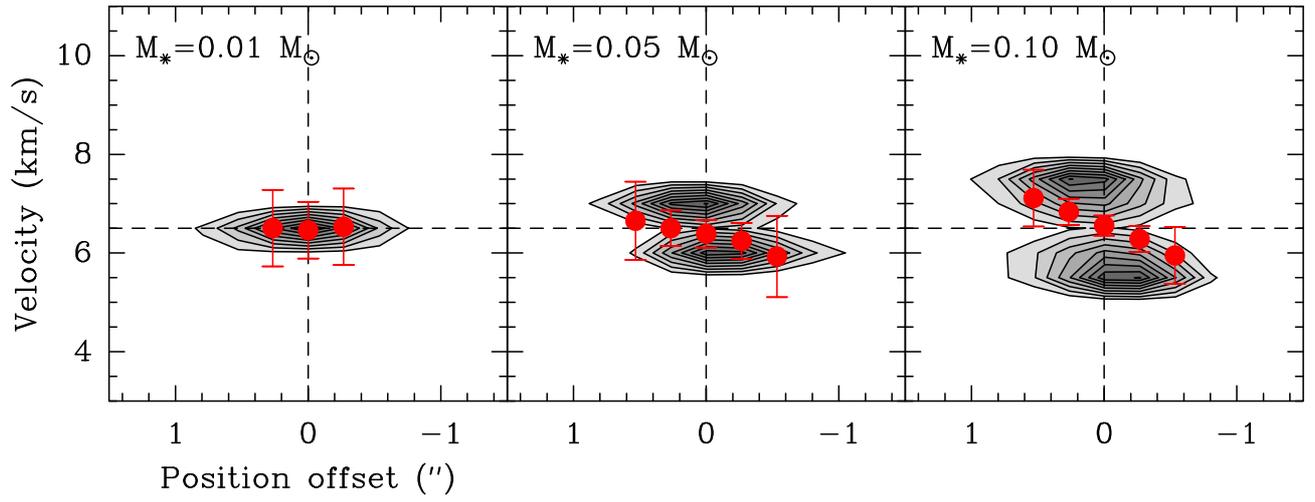}
  \caption{Synthetic position-velocity diagrams for the 1~mm methanol
    line (gray contours). The contour levels and axis ranges are the same as in
    Fig.~\ref{fig:pv}. The red points with error bars show the line
    centroid along the cut. The diagrams have been computed assuming
    that the emission originates in a Keplerian disk with different
    masses of the central protostar (see text).}
  \label{fig:synthetic-pv}
\end{figure*}

To test the Keplerian disk scenario, we have computed synthetic
P.V. diagrams for the 1~mm methanol line. We have assumed that the
emission follows a Gaussian distribution, with a FWHM of
0.44\arcsec~(as determined from the visibility fit). The line-of-sight
velocity is assumed to be given by \citep[see, e.g.,][]{Guilloteau12}

\begin{equation}
  v\left(r,\theta\right) =  \mathrm{sin}\left(i\right) \, \mathrm{cos}\left(\theta\right) \,
  v_\mathrm{kep}\left(r\right) + v_\mathrm{LSR}
  \label{eq:1}
\end{equation}

\noindent
where $r$ and $\theta$ are the cylindrical coordinates in the disk
plane, $i$ is the inclination of the disk rotation axis with respect
to the line of sight (i.e., $i = 90\degr$ for an edge-on disk),
$v_\mathrm{LSR}$ is the source velocity in the local standard of rest,
and $v_\mathrm{kep}\left(r\right)$ is the Keplerian velocity

\begin{equation}
  v_\mathrm{kep}\left(r\right) = \sqrt{\frac{G \, M_{*}}{r}}
\end{equation}

\noindent
with $G$ the gravitational constant and $M_*$ the mass of the central
protostar. The line is assumed to have a Gaussian shape, with a local
linewidth $\Delta v\left(r\right)$ ($\mathrm{FWHM}/\sqrt{8 \,
  \mathrm{ln} \left( 2 \right)}$) equal to $0.1 \,
v_\mathrm{kep}\left(r\right)$, in agreement with the value determined
in DM Tau by \cite{Guilloteau12}. Although this value is uncertain, it
cannot exceed $v_\mathrm{kep}\left(r\right)$ because a disk with a
larger $\Delta v\left(r\right)$ would be unstable. We assume that the
disk is rotating about an axis with $i = 60\degr$. This is consistent
with the disk rotating about the outflow axis, which has $i \ge
45\degr$ \citep{Codella13}. Finally, we assume a $v_\mathrm{LSR}$ of
6.5~km/s, a source distance of 235~pc, and $M_*$ is left as a free
parameter. The disk is not truncated in radius; we merely assume that
its line emission decreases as a Gaussian, as mentioned above. We
compute a synthetic data cube with a pixel size of 0.1$\arcsec$, and a
channel width of 0.5~km/s. The intensity of the emission as a function
of channel is computed assuming that the line-of-sight velocity varies
as a function of $r$ and $\theta$ following Eq.~(\ref{eq:1}). This
data cube is then convolved with the synthesized beam, assumed to be a
Gaussian with a FWHM of 0.8\arcsec. The convolved data cube is then
scaled so that the peak intensity matches the observations
($453~\mathrm{mJy/beam}$). Finally, Gaussian noise is added in each
channel so that the signal-to-noise ratio also corresponds to the
observations ($\sim 21$).

Figure~\ref{fig:synthetic-pv} shows the synthetic position velocity
diagrams we obtain for different values of $M_*$. We find that our
observations are inconsistent with a Keplerian disk, regardless of the
mass of the central object. For $M_* = 0.05~\mathrm{M_\sun}$ and
$0.1~\mathrm{M_\sun}$, the synthetic P.V. diagrams have two peaks,
while the observed P.V. diagram has one peak (see
Fig.~\ref{fig:pv}). For $M_* = 0.01~\mathrm{M_\sun}$, the synthetic
P.V. diagram is single peaked, and the predicted first-order moment
along the cut is constant, within the error bars. However, the
predicted linewidth is much smaller than observed (compare the size of
the P.V. contours along the vertical axis in Fig.~\ref{fig:pv}
and~\ref{fig:synthetic-pv}). A broader linewidth would require a
larger value of $\Delta v\left(r\right)$. However, we find that even
for a very turbulent disk with $\Delta v\left(r\right) = 0.9 \,
v_\mathrm{kep}\left(r\right)$, the predicted linewidth is 2-3 times
smaller than observed.  We conclude that the observed methanol
emission can not arise from a Keplerian disk, and must originate in
the infalling and perhaps slowly rotating inner envelope. A more
detailed modeling of the methanol line emission is needed to determine
the precise velocity field of the inner region of the protostar
envelope.

\end{appendix}

\end{document}